\documentclass[conference]{IEEEtran}
\IEEEoverridecommandlockouts

\usepackage{cite}
\usepackage{amsmath,amssymb,amsfonts}
\usepackage[hidelinks,urlcolor=blue]{hyperref} 
\usepackage{graphicx}
\usepackage{textcomp}
\usepackage{xcolor}
\usepackage{gensymb}
\usepackage{algorithm}
\usepackage{algorithmic}
\graphicspath{{figures/}}
\def\BibTeX{{\rm B\kern-.05em{\sc i\kern-.025em b}\kern-.08em
    T\kern-.1667em\lower.7ex\hbox{E}\kern-.125emX}}

\newcommand{\cng}[1]{\textcolor{black}{#1}}

\begin{document}

\title{Bias Variation Compensation in Perimeter-Gated SPAD TRNGs

\thanks{This article is based on work supported by the National Science Foundation under Grant No. 2442346. \textcolor{red}{“This is the accepted version of this work at the IEEE MWSCAS 2025 Conference and submitted to the IEEE for possible publication. Copyright may be transferred without notice, after which this version may no longer be accessible.”}}
}

\author{\IEEEauthorblockN{Md Sakibur Sajal, Hunter Guthrie, and Marc Dandin}
{Department of Electrical and Computer Engineering},\\
{Carnegie Mellon University,}
{Pittsburgh, Pennsylvania, 15213, USA}\\
\\
{email: mdandin@andrew.cmu.edu}   }


\maketitle

\begin{abstract}
Random number generators that utilize arrays of entropy source elements suffer from bias variation (BV). Despite the availability of efficient debiasing algorithms, optimized implementations of hardware friendly options depend on the bit bias in the raw bit streams and cannot accommodate a wide BV. In this work, we present a $\mathbf{64\times64}$ array of perimeter gated single photon avalanche diodes (pgSPADs), fabricated in a $\mathbf{0.35~\mu m}$ standard CMOS technology, as a source of entropy to generate random binary strings with a BV compensation technique. By applying proper gate voltages based on the devices' native dark count rates, we demonstrate less than $\mathbf{1\%}$ BV for a raw-bit generation rate of $\mathbf{2~kHz/pixel}$ at room temperature. The raw bits were debiased using the classical iterative Von Neumann's algorithm and the debiased bits were found to pass all of the 16 tests from NIST's Statistical Test Suite.
\end{abstract}


\section{Introduction}
Random number generators (RNGs) find applications in various domains such as cryptography, gaming, scientific research, computer simulations, and machine learning~\cite{Bikos2023RandomApplications,Lugrin2023,Seyhan2022ClassificationTaxonomy}. In secure communication systems, they are an indispensable building block for generating nonces and session keys~\cite{Rajski2023AApplications}. Depending on the application's need, they can be implemented using natural sources of randomness, \textit{i.e.}, as true RNGs (TRNGs)~\cite{Yao2023DCDRO:AOscillator,Woo2024TrueLaCoO3} or using complex, yet deterministic algorithms, \textit{i.e.}, as pseudo-RNGs (PRNGs)~\cite{Bhattacharjee2022AStudies,Min2013StudyApplication}. However, TRNGs are preferred over PRNGs when unpredictability is of the utmost concern~\cite{prngnotrobust}.

\begin{figure}
    \centering
    \includegraphics[width=\linewidth]{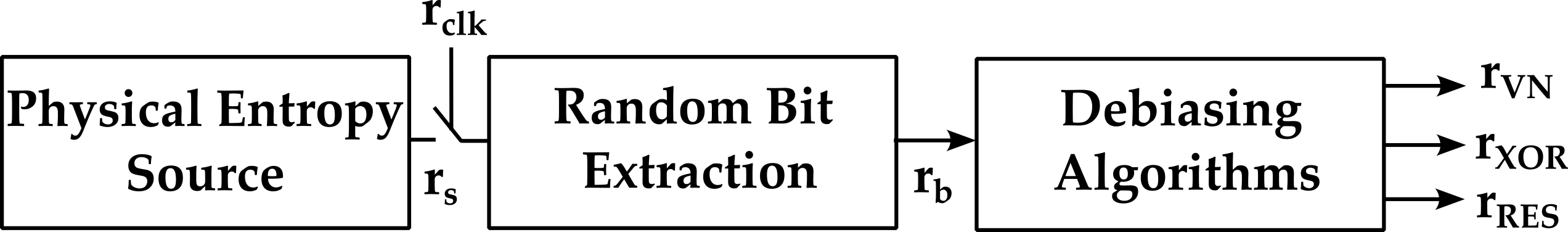}
    \vspace{-15pt}
    \caption{Basic building blocks of a TRNG including an entropy source fluctuating between random states at an average rate of $r_s$. The states are sampled periodically ($r_{clk}$) to generate one or more bits ($r_b$) depending on the bit extraction method. The raw bit sequences may need to be debiased to generate one or more processed bit-streams (\textit{i.e.}, $r_{VN},r_{XOR},r_{RES}$). Example is shown for the iterative Von Neumann (IVN) debiasing algorithm.}
    \vspace{-5pt}
    \label{fig:trng}
\end{figure}

As shown in Fig.~\ref{fig:trng}, a TRNG requires a physical entropy source, which is sampled for its fluctuating random states over time by a bit extraction mechanism~\cite{Chernov2018TowardsDesign}. Depending on the extraction criteria, the raw bits may require post-processing by one or more debiasing stages in order to increase the entropy in the output bit-streams~\cite{Balasubramanian2018HashApplications,dichtlref}.

TRNGs can be constructed using any classical source of noise, such as the thermal noise of resistors~\cite{Gong2019TrueNoise}. However, quantum RNGs (QRNGs) are superior in terms of security, which leverage quantum randomness~\cite{Jacak2021QuantumNumbers}. For example, the uncertainty pertaining to photon statistics can be utilized to generate quantum random numbers~\cite{Burri2014SPADsBeyond,Incoronato2021LinearGenerator,Tontini2019SPAD-BasedFPGA}. As such, single photon avalanche diodes (SPADs) are commonly found to be the photon transducers in photon-based QRNGs.  

SPADs are p-n junction diodes that are operated beyond their nominal reverse breakdown voltage; this operational regime is commonly known as the Geiger mode. The resulting high electric field at the depleted junction makes the devices sensitive enough to detect individual incident photons. An electron-hole pair generated from photon absorption can probabilistically initiate an avalanche event which can be time-tagged by an external sensing circuit \textit{i.e.}, using a time-to-digital converter (TDC)~\cite{Tontini2019SPAD-BasedFPGA,Yan2015High-SpeedPhotons}. Similarly, the number of avalanche events within a defined time window can be accumulated by integration~\cite{Acerbi2018AStructure,Massari2022AApplication}. Random bits can be generated from these random detection times or random amounts of detected events.

SPAD arrays provide increased throughput. They are usually coupled to a photon source attenuated with neutral density filters~\cite{Nicola2017APhotons}. The latter feature increases hardware overhead. These systems are also limited by the \cng{nonuniform illumination from the photon source leading to bias variation across the array. Moreover, device manufacturing variation can further cause bias variation, posing non-trivial challenges}\cite{Massari2015AGenerator,Massari201616.3Temperature}. 

An alternative to the photon-based systems is SPAD dark noise-based TRNGs~\cite{Lin2019TruePhotomultiplier,Tawfeeq2009ACounts}. Recently, we have demonstrated a perimeter gated SPAD (pgSPAD) dark noise-based TRNG with near zero bias and temperature stability~\cite{Sajal2024TrueDiode}. This variant of SPADs utilizes a polysilicon gate over the perimeter junction to modulate the dark carrier generation rate~\cite{Sajal2022Perimeter-GatedProbability}. In this work, we extend our study to analyze the behavior of a pgSPAD array in terms of its native BV and the underlying trade-offs with bit generation rate. In particular, we show that hardware-friendly algorithms like iterative Von Neumann (IVN) debiasing~\cite{Rozic2016IteratingConstraints} can benefit from BV reduction using perimeter gating, which obviates the need for custom optimization.

\section{pgSPAD Array: The Entropy Source}
\begin{figure}
    \centering
    \includegraphics[width=\linewidth]{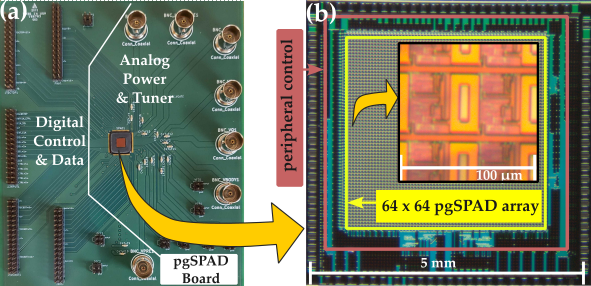}
    \vspace{-15pt}
    \caption{A pgSPAD die-on-PCB prototype board (a) with the digital (control $\&$ data) and the analog (power $\&$ tuner) ports. The $5~mm\times5~mm$ die (b) contains a $64\times64$ pgSPAD pixel array, surrounded by the peripheral control blocks for pixel selection, active quench $\&$ reset, and data readout. Inset shows a photo-micrograph of the $50~\mu m \times50~\mu m$ pixels, each containing a pgSPAD and its sensing circuitry. In-depth circuit details can be found in Ref.~\cite{Sajal2022Perimeter-GatedProbability}.}
    \label{fig:pcb}
\end{figure}

Figure~\ref{fig:pcb} (a) shows a prototype die-on-PCB with a $5~mm\times5~mm$ pgSPAD array fabricated in a $0.35~\mu m$ standard CMOS process. A photomicrograph (Fig.~\ref{fig:pcb} (b)) of the die shows the $64\times64$ pixel array, surrounded by the control logic blocks. The pixel architecture, consisting of a diode and its sensing circuitry, is shown in the inset. In-depth circuit details and operational characteristics of the device can be found in ref.~\cite{Sajal2022Perimeter-GatedProbability}. Nevertheless, we briefly describe the operation of the device below.

\begin{figure}
    \centering
    \includegraphics[width=\linewidth]{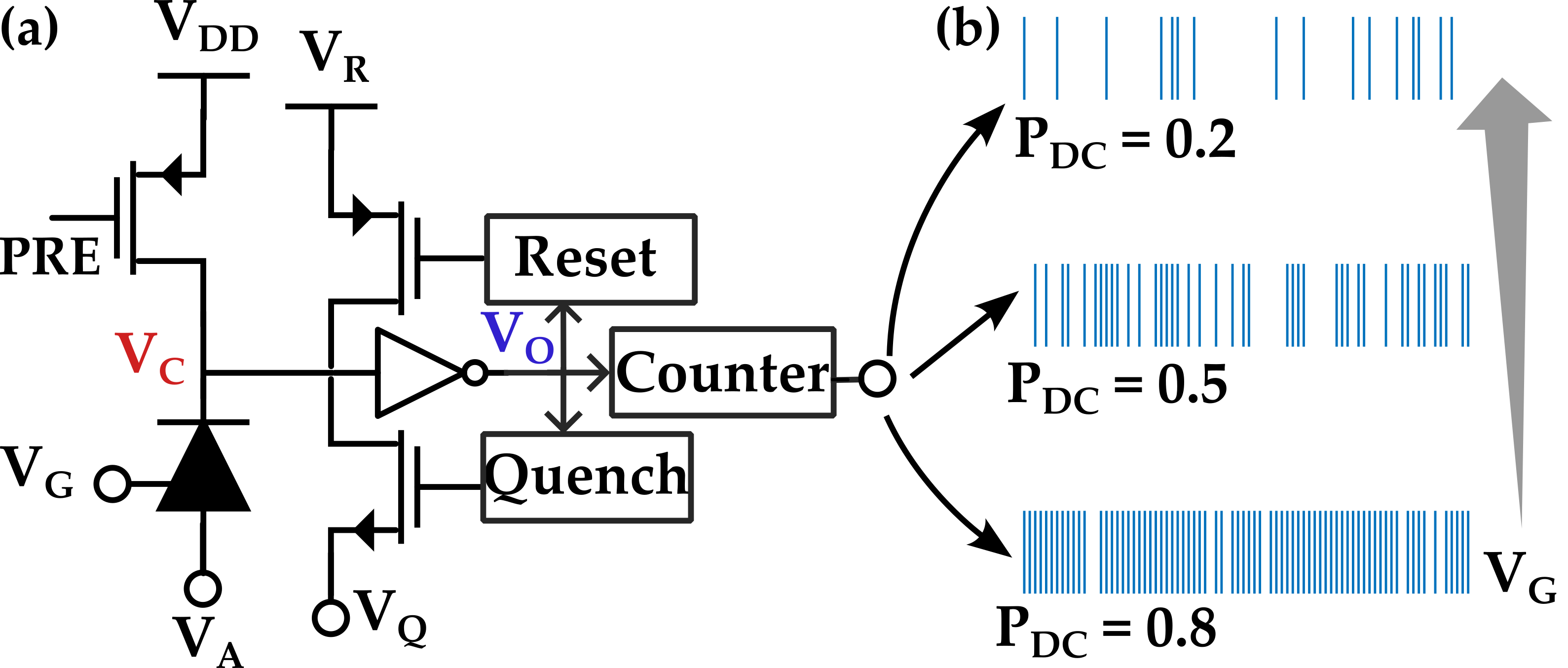}
    \vspace{-15pt}
    \caption{Simplified schematic of the dark count based TRNG using a perimeter gated SPAD (a). Analog voltage of the cathode ($V_C$) is digitized at the inverter's output ($V_o$) which triggers the active quench and reset, and the counter blocks. The dark count probability ($P_{DC}$) can be altered by the perimeter gate voltage ($V_G$), generating bit-steams with different biases (b).} 
    \label{fig:sch}
\end{figure}

Figure~\ref{fig:sch} (a) shows a simplified schematic of a generic pixel. When a particular pixel is selected, a precharge ($PRE$) signal pulls the cathode voltage, $V_C$ to $V_{DD}$ while the counter gets reset. By de-asserting $PRE$, the device is allowed to be triggered by photo- or dark-carriers while being actively quenched to $V_Q$ and reset to $V_R$ in a free-running mode. Specifically in the dark, the average triggering rate, $\lambda_{0}$, also known as the dark count rate (DCR) is proportional to the dark carrier generation rate, dominated by thermal excitation, trap assisted tunneling, and band-to-band tunneling (B2B)~\cite{Xu2017ComprehensiveTechnologies}. 

B2B, and consequently the DCR, can be modulated by the excess bias voltage, $V_{ex} = (V_R-V_A)-V_{brk}$ and the perimeter gate voltage, $V_G$. Here, $V_A$ is the applied anode voltage and $V_{brk}$ is the diode's breakdown voltage.

Considering afterpulsing and deadtime, the probability of observing a dark event ($P_{DC}$) within $T_{int}$ is expressed as
\begin{equation}
\label{eq:pdc}
    P_{DC} = 1 -e^{-\lambda_{eff}T_{int}}
\end{equation}
where, 
\begin{equation}
    \lambda_{eff} = \frac{\lambda_0}{1-\tilde{p}_a + \lambda_0(\tau_R/T_{int})}.
\end{equation}
Here, $\tilde{p}_a$ is the adjusted afterpulsing probability, and $\tau_R$ is the adjusted device recovery time taking quenching and probabilistic reset time into consideration~\cite{Straka2020CountingIllumination}.

Figure~\ref{fig:sch} (b) shows the time-series of the LSB flip of the counter as $P_{DC}$ is varied via the gate voltage ($V_G$). Each spike denotes a dark event detection. As $V_G$ increases, the probability of a dark event detection decreases~\cite{Sajal2024TrueDiode}.

\section{Random Bit Extraction Protocol}

\begin{figure}
    \centering
    \includegraphics[width=0.95\linewidth]{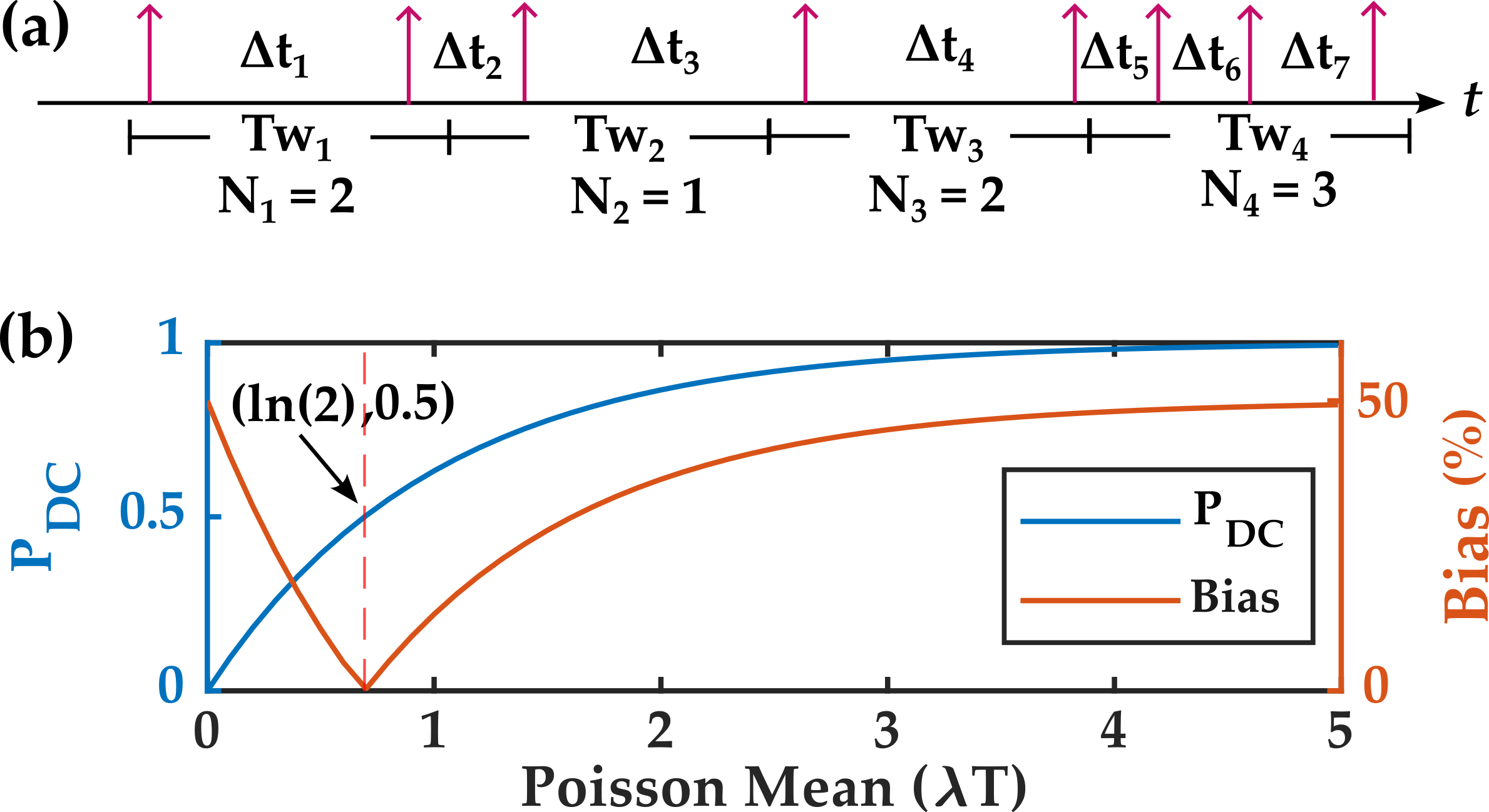}
    \vspace{-5pt}
    \caption{Conventional bit generation protocols from interdetection times and the number of detected events (a). Zero bias condition can be achieved (b) for equal probability of $N_i=0$ and $N_i\neq0$ by maintaining $\lambda T = ln(2)$.}
    \label{fig:mtd}
\end{figure}

Comparison between the inter-detection times $\Delta t_i$' s (see Fig.~\ref{fig:mtd} (a)) usually involves power-consuming TDCs. An energy-efficient alternative to this is the counter-based approach to track the number of detected events, $N_i$'s in each fixed time-window, $Tw_i$.
By properly adjusting the activity rate, $\lambda$ and the window duration, $T$, it is possible to operate the device at zero bias with equal probability of $N=0$ and $N\neq0$ (see Fig.~\ref{fig:mtd} (b)). Specifically, it only requires that
\begin{equation}
\label{eq:ln2}
    \lambda T = ln(2).
\end{equation}
\cng{With bias, $b$ defined as $b = \frac{|p_1-p_0|}{2} = \frac{|P_{DC}-(1-P_{DC})|}{2}$,
where $p_1$ and $p_0$ are the probabilities of $N=0$ and $N\neq0$, respectively, Eqs.~\ref{eq:pdc} and~\ref{eq:ln2} ensure that $b\rightarrow0$ for different combinations of dark activities and integration times.}

However, $T$ is defined and fixed for all pixels in an array by the bit-sampling rate. Hence, the intrinsic variation of $\lambda$ leads to a pixel-wise bias resulting in
\begin{equation}
\label{eq:bbv}
    b_i = |0.5-e^{-\lambda_{eff,i}T_{int}}|\times100\%.
\end{equation}
A higher degree of bias degrades the overall entropy of the output. Therefore, as a standard practice, the raw bits are post processed by debiasing stages such as the IVN algorithm to restore the entropy in the output streams.

\section{Debiasing Method: Iterative VN Algorithm}

Von Neumann (VN) post processing is a hardware-friendly method, famous for debiasing raw bit sequences in a resource-constrained setup~\cite{vonN51}. While this method guarantees full entropy recovery, provided that the bits are independent, it loses a substantial amount of bits in the process~\cite{Burri2014SPADsBeyond}. 

\begin{table}
    \caption{Bit Assignment Rules for the IVN Debiasing Algorithm}
    \label{tab:rules}
    \centering
    \begin{tabular}{|cc|c|c|c|} \hline
         \multicolumn{2}{|c|}{$S_{PAIR}$}& $S_{VN}$ & $S_{XOR}$ &$ S_{RES}$  \\ \hline
         0& 0 & - & 0 & 0 \\
         0& 1 & 0 &  1&  -\\
         1& 0 & 1 &  1&  -\\
         1& 1 & - &  0&  1\\ \hline
         \multicolumn{2}{|c|}{Bias, $b$}& 0  & $2\cdot b^2$  & $\frac{2\cdot b}{1+ 4\cdot b^2}$ \\ \hline
         \multicolumn{2}{|c|}{Rate, $r$}& $(0.25-b^2)\cdot r$ & $0.5\cdot r$ & $(0.25+b^2)\cdot r$ \\ \hline
    \end{tabular}
\end{table}

Table~\ref{tab:rules} shows the bit assignment rules for the VN process. The raw bits in a sequence, $S$ are paired up to create $S_{PAIR}$ which are either discarded or replaced by a $0$ or a $1$ depending on the pair combinations shown in the $S_{PAIR} \rightarrow S_{VN}$ transformation. This results in a $75\%$ reduction in the throughput at the very least depending on the bias in the input string~\cite{Rozic2016IteratingConstraints}.

To increase the throughput, an iterative VN (IVN) process was proposed by Peres with the idea of re-applying VN on the discarded information~\cite{Peres1992IteratingBits}. In theory, it is possible to extract the maximum entropy available from the original bit sequence by iterating the IVN procedure to infinity on the resulting 
XOR sequence, $S_{XOR}$ and the residual sequence, $S_{RES}$ after each iteration according to the rules shown in Table~\ref{tab:rules} ~\cite{Rozic2016IteratingConstraints}. 

\begin{figure}
    \centering
    \includegraphics[width=0.95\linewidth]{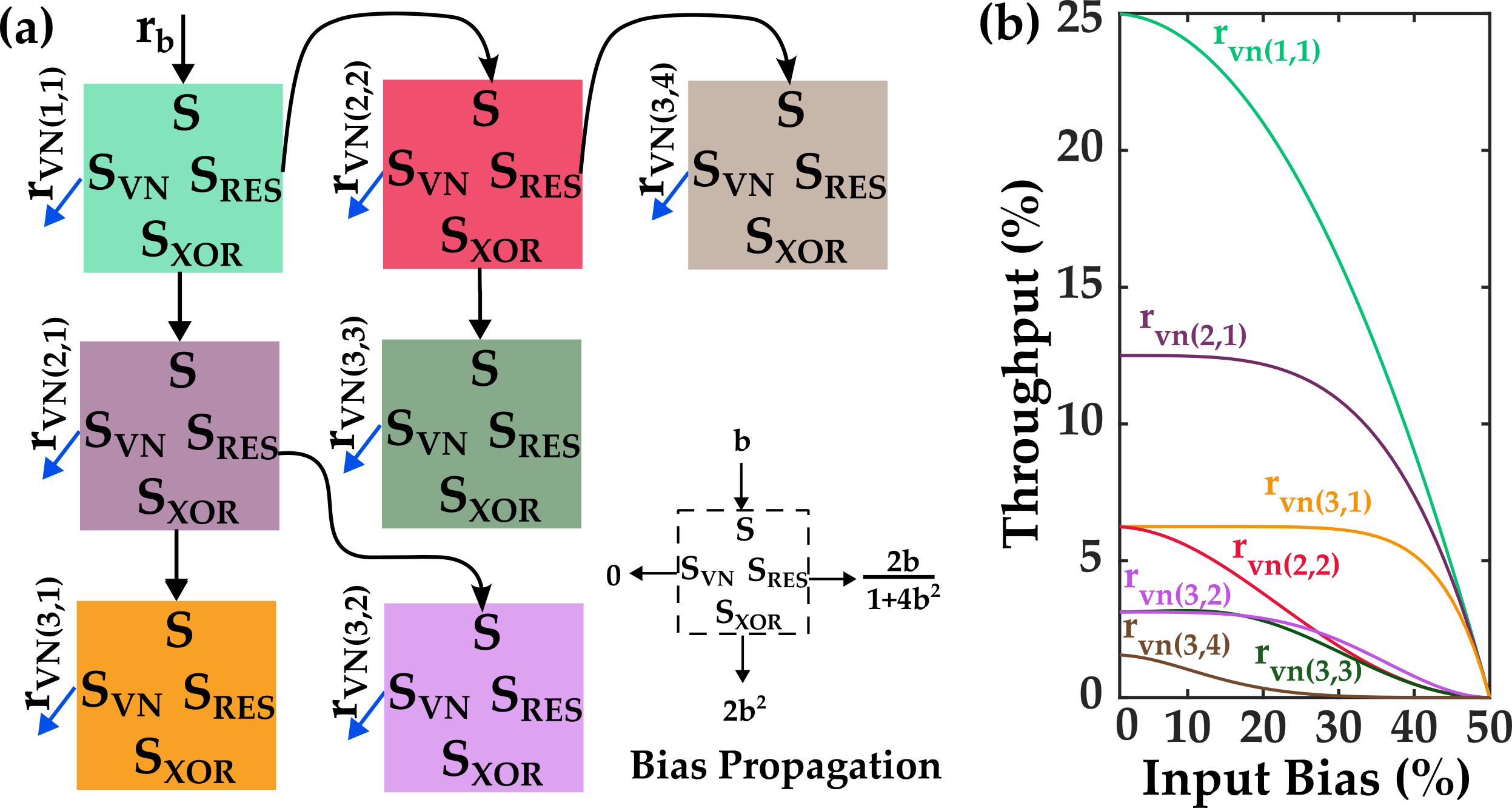}
    \vspace{-5pt}
    \caption{Classical implementation of IVN~\cite{Rozic2016IteratingConstraints} (a) with 7 elements. Throughput distribution and variation with respect to the bias in the input string (b). More than a $12\%$ bias reduces the total throughput by more than $5\%$.}
    \label{fig:ivn}
\end{figure}

However, due to area and energy constraints, a finite number of stages are utilized as shown by the block diagram in Fig.~\ref{fig:ivn} (a) depicting a classical IVN implementation. Since each stage takes up the same area and power for a rapidly diminishing throughput as shown in Fig.~\ref{fig:ivn} (b), a procedure to optimize the connection-tree for a given number of stages was proposed in Ref.~\cite{Rozic2016IteratingConstraints}. However, such optimization methods depend on the bias, $b$ in the raw input since both the throughput and the bias propagation directly depend on it (see Table~\ref{tab:rules} and Fig.~\ref{fig:ivn} (b)). Additionally, it is impractical to implement a custom optimized connection-tree for each TRNG unit in an array to accommodate a wide bias variation (BV)~\cite{Massari201616.3Temperature}. 

\section{Experimental Results and Discussions}
\subsection{Native DCR and the Bias Variation Across the Array} 
\begin{figure}
    \centering
    \includegraphics[width=\linewidth]{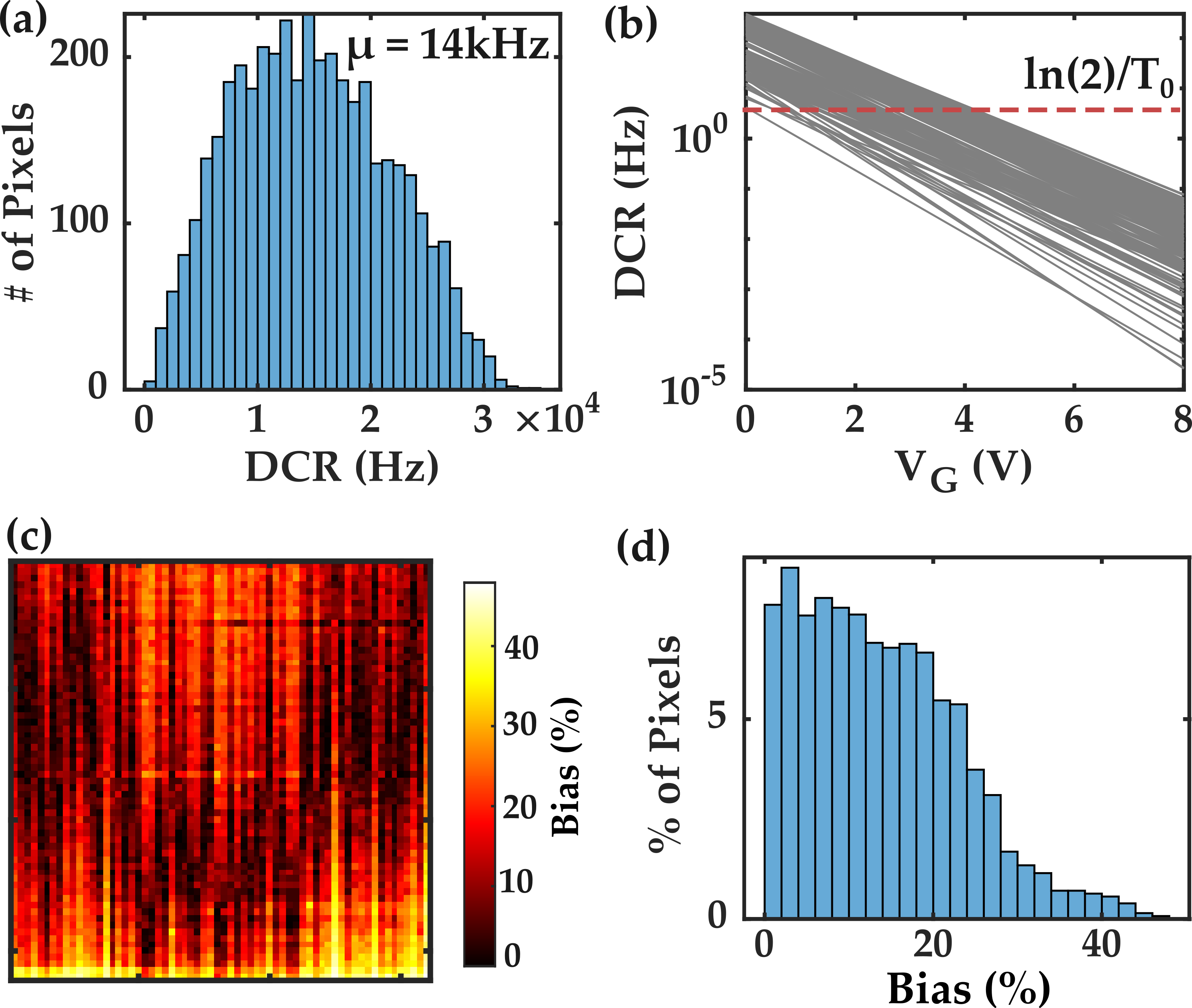}
    \vspace{-15pt}
    \caption{DCR distribution (a) at room temperature for an excess bias of $\sim 2~V$ and a perimeter gate voltage of $0~V$. Exponential decrease of individual DCR with the increase of the gate voltage magnitude (b). Spatial distribution (c) and the histogram (d) of bias for $0~V$ on the gate with $T = ln(2)/\bar{\lambda_0} \sim 50~\mu s$.}
    \label{fig:dcr}
\end{figure}

Figure~\ref{fig:dcr} (a) shows the DCR histogram for $4,096$ pixels operating at room temperature with $V_{ex} \sim 2~V$ and $V_G = 0~V$. We see a DCR range of $\sim 35~kHz$ with an average, $\bar{\lambda_0} \sim14~kHz$. Figure~\ref{fig:dcr} (b) shows the exponential decrease of the DCR across the entire array with the increase of the perimeter gate voltage magnitude. Here, we have presented the fitted trend lines for 100 randomly selected pixels, including the most active and the least active ones. We can empirically express the DCR, $\lambda_i$ of a unit as the function of $V_{G,i}$ as
\begin{equation}
    \lambda_i(V_G) = \lambda_{0,i}e^{-\alpha_i V_{G,i}}
\end{equation}
where, $\alpha_i (V^{-1})$ is the gate voltage coefficient of the exponent.

Therefore, it is possible to apply pixel-specific optimal gate voltage $V_{G,i}^*$ to make the array exhibit the same DCR, \textit{i.e.}, $ln(2)/T_0$ (shown by the red broken line) by all the pixels. Here, $T_0$ is the sampling time for a $0$ bias variation.

Figures~\ref{fig:dcr} (c) and (d) show the spatial distribution and the histogram of the native bias of the array, respectively, for $V_G = 0~V$ and $T = ln(2)/\bar{\lambda_0} \sim 50~\mu s$. This corresponds to a raw bit generation rate of $20~kHz/pixel$.
The array average bias was found to be $\sim 14\%$ with a BV $\sim9\%$, calculated as the standard deviation of the bias across the array.

\subsection{Bias Variation Compensation with Perimeter Gating}

By increasing the gate voltage magnitude, it is possible to reduce the dark activity of each pixel to a desired level (see Fig.~\ref{fig:dcr} (b)). However, pixels exhibiting DCRs below that desired activity level will still contribute to the bias variation. Therefore, it is necessary to select a DCR activity level that is close to the array minimum to achieve a $\sim0$ BV.

\begin{figure}
    \centering
    \includegraphics[width=\linewidth]{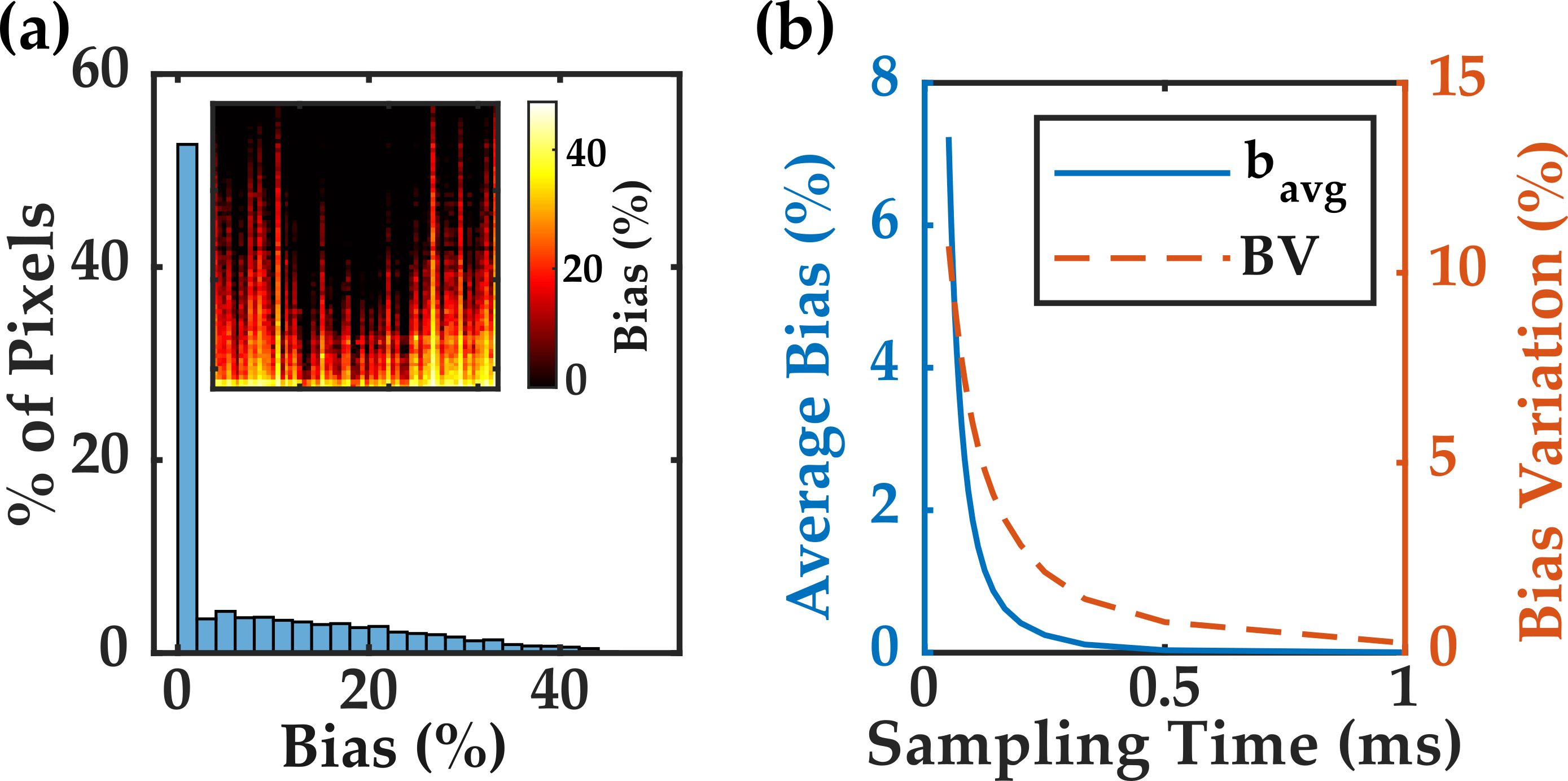}
    \vspace{-15pt}
    \caption{Reduction of average bias from $~14\%$ to $~8\%$ by applying appropriate gate voltages to pixels exhibiting higher than the native average DCR (a). Relationship of the average bias and the bias variation with the bit-sampling time (b) showing the trade-off between bias and bit-generation rate. We found $T = 0.5~ms$ corresponding to a BV $<1\%$.}
    \vspace{-10pt}
    \label{fig:bbvred}
\end{figure}

For instance, Fig.~\ref{fig:bbvred} (a) shows the histogram and the spatial variation (inset) of the bias as we reduce the activity of the pixels exhibiting higher DCR than the native average, $\bar{\lambda_0} = 14~kHz$ by applying appropriate gate voltages. Although this reduced the array average bias, $b_{avg}$ to $\sim8\%$ from its native value of $\sim14\%$, BV is still $\sim9\%$ since $50\%$ of the pixels have lower dark count activity than the one required to satisfy $\lambda T = ln(2)$ for $T=50~\mu s$. Therefore, to reduce the required $\lambda$, we needed to increase the sampling time, $T$.

Figure.~\ref{fig:bbvred} (b) shows the $b_{avg}$ and the BV as we swept the sampling time from $0.1~ms$ to $1~ms$. For each value of $T$, the gate voltages were adjusted to confine the dark noise activity of individual pixels close to the required value, \textit{i.e.}, $ln(2)/T$.

Hence, we readily see the trade-off between the sampling time or bit-generation rate with the BV as both the average bias and the bias variation decrease as we increase the sampling time. In other words, a lower bit generation rate allows for a lower average bias and a near zero BV. Consequently, we selected a raw bit-generation rate of $2~kHz/pixel$  to achieve a BV less than $1\%$ to use the classical IVN connection-tree without needing any optimization. However, it should be noted that higher bit-generation rates can be accommodated while incurring a greater BV if optimized IVN connection-trees with more stages are deployed~\cite{Rozic2016IteratingConstraints}. 

\subsection{Bit-Streams Generation and Testing with the NIST STS}
We generated random bits from random pixels across the array with the target maximum BV of $<1\%$. Each of the raw bits were debiased using a 3-stage classical IVN algorithm to obtain $>1M$bits/string. The resulting bit strings were tested using the 16 statistical tests prescribed by NIST ~\cite{niststs}.

\begin{figure}
    \centering
    \includegraphics[width=0.8\linewidth]{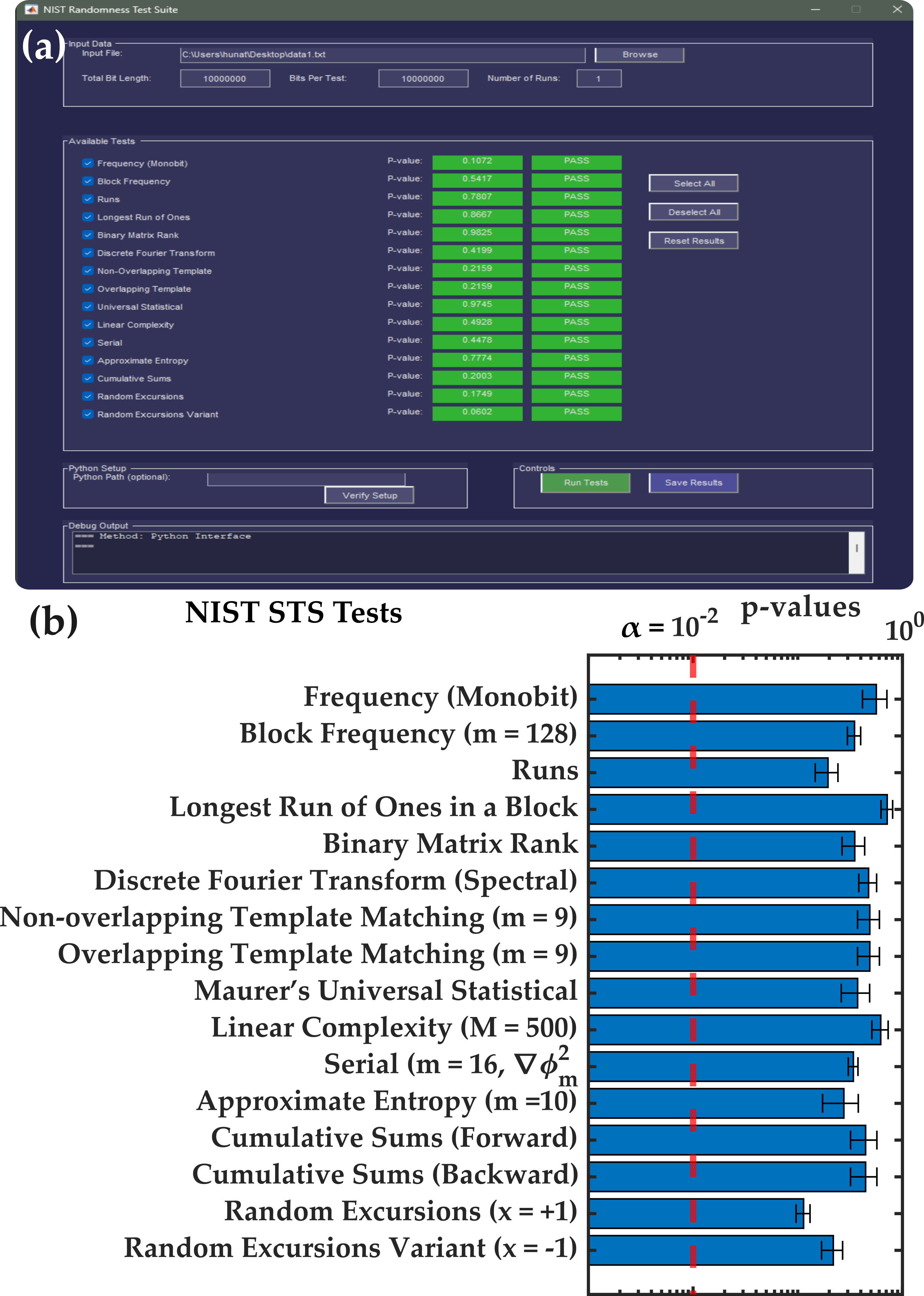}
    \vspace{-5pt}
    \caption{MATLAB GUI~\cite{nistgui} created to test the quality of the debiased bit-strings (a) and the p-value bar plots (b) by administering the 16 statistical tests prescribed by NIST~\cite{niststs}.}
    
    \label{fig:gui}
\end{figure}

We have created a MATLAB based GUI for the NIST Statistical Test Suite (STS) as a part of this work, which is easy to use and flexible on the input data file type. Fig.~\ref{fig:gui} (a) shows the GUI with the results of a random bit string passing all the tests similar to the rest of the debiased bit-strings.

Fig.~\ref{fig:gui} (b) shows the tests that were performed along with the p-value ranges obtained from the tests. Here, randomness is confirmed when a p-value is greater than the confidence level of a test, which we set to the default value of $\alpha = 0.01$. 

All tests were performed with the default setup prescribed by NIST.  Furthermore, we note that for the random excursions and random excursions variant tests, we considered the minimum of the possible 8 and 18 p-values, respectively, to conclude a pass or fail for the given string. 

\section{Conclusions}
In this work, we have presented the utility of the perimeter gating technique to reduce the bias variation (BV) in a true random number generator (TRNG) implemented with a single photon avalanche diode (SPAD) array. Our dark noise-based approach eliminates the need for bulky optical components such as a light source, neutral density filter, lenses, and diffusers. By keeping the BV less than $1\%$, we were able to use the classical iterative Von Neumann debiasing without needing to implement any custom optimization for the connection-tree on a pixel-by-pixel basis. This proposed pgSPAD TRNG array can be easily integrated in an area-constrained application, such as low-power remote IoT devices needing frequent authentication to establish secure communication links. Furthermore, the TRNG can be augmented by any pseudo-RNG to increase the bit-generation rate impacted by lower BV operating conditions. \cng{For larger array formats, sub-arrays can share the the perimeter gate control voltage.}


\bibliographystyle{IEEEtranDOI}
\bibliography{combined_bibfile}

\end{document}